\documentclass[conference]{IEEEtran}
\usepackage{graphicx}
\usepackage{float} 
\usepackage{url} 
\usepackage{color}

  \usepackage{caption}
  \usepackage{subcaption}
  \usepackage{setspace}

\newenvironment{noindlist}
 {\begin{list}{\labelitemi}{\leftmargin=1.2em \itemindent=-.5em}}
 {\end{list}}

\ifCLASSINFOpdf
\else
\fi
\begin{document}
%
\title{Improve the Sustainability of Internet of Things Through Trading-based Value Creation }


\author{\IEEEauthorblockN{Charith Perera\IEEEauthorrefmark{1}\IEEEauthorrefmark{2} and 
 Arkady Zaslavsky\IEEEauthorrefmark{2}
}
\IEEEauthorblockA{\IEEEauthorrefmark{1}Research School of Computer Science, The Australian National University, Canberra, ACT 0200, Australia}
\IEEEauthorblockA{\IEEEauthorrefmark{2}CSIRO Computational Informatics, Canberra, ACT 2601, Australia}}


%


\maketitle

\begin{abstract}
Internet of Things (IoT) has been widely discussed over the past few years in technology point of view. However, the social aspects of IoT are seldom studied to date. In this paper, we discuss the IoT in social point of view. Specifically, we examine the strategies to increase the adoption of IoT in a sustainable manner. Such discussion is essential in today's context where adoption of IoT solutions by non-technical community is slow. Specially, large number of IoT solutions making their way into the market every day. We propose an trading-based value creation model based on sensing as a service paradigm in order to fuel the adoption of IoT. We discuss the value creation and its impact towards the society especially to households and their occupants. We also present results of two different surveys we conducted in order to examine the potential acceptance of the proposed model among the general public.

\end{abstract}


%
\IEEEpeerreviewmaketitle

\section{Introduction}
\label{sec:Introduction}

Internet of Things (IoT) has been a popular term over the past few years.  The main reasons behind such interest are the capabilities and sophistication that the IoT will bring to the society \cite{P003}. It promises to create a world where all the objects around us are connected to the Internet and communicate with each other with minimum human intervention. The ultimate goal is to create \textit{`a better world for human beings'}, where objects around us know what we like, what we want, and what we need and act accordingly without explicit instructions \cite{P040}. The Internet of Things allows \textit{``people and things to be connected Anytime, Anyplace, with Anything and Anyone, ideally using Any path/network and Any service''} \cite{P029}. During the past decade, the IoT has gained significant attention in academia as well as industry. The interest is also fuelled by both predictions and estimations \cite{ZMP003}. Large number of solutions have been introduced to the IoT market place by different types of organization ranging from start-ups, academic institutions to large enterprises \cite{P596}. 

Despite the strong interest, the adoption of IoT solutions among the general public (i.e. non-technical community and typical households) is still remaining low. Our objective is to address this issue of low adoption. The research questions we address in this paper are: \textit{1) how to increase the adoption of IoT solutions among non-technical community?} and  \textit{2) how to make the adoption more sustainable by motivating the consumers in typical households?}. In typical wireless sensor network domain as well as in IoT domain, sensors are deployed in order to perform a particular task. However, in IoT, once sensors are deployed, data they generate are used to accomplish multiple tasks and objectives including the tasks that are never even think of at the time of deployments. The proposed model is expected to strengthen  such re-usability and resource optimization where it will also stimulates collaborative sensing.

We propose an trading-based value creation model to motivate the consumers to deploy IoT solutions at their households. It allows to sell the data generated by those smart solutions in an action-based IoT market place. At the same-time, parties who are interested in accessing such data can bid and get access to them. Such a model is  for all the parties who are. There are many advantages of sharing deployed sensors and the data they generate. Some of the significant  benefits of the proposed model are reduction of data acquisition cost, collect  previously unavailable data, real-time data for decision making and policy making, innovations and novel applications. The social impact of these benefits is  later.

The remainder of this paper is organised as follows: we briefly explain the sensing as a service model and its contribution towards the sustainability of IoT in Section \ref{sec:Sensing_as_a_Service Model}. In Section \ref{sec:The Future}, we explain how trading-based IoT market place work, using an example scenario. The major advantages, value creation, and the impact of the proposed model towards society are discussed in Section \ref{sec:Value_Creation}. This section glues all the concepts together in order to highlight the impact.\textcolor{black}{In Section \ref{sec:Discussion}, we present the results of two surveys that highlights the community responses to the proposed model which provides valuable insights.} Final Section presents the concluding remarks.

\section{Trading-based Sensing as a Service Model}
\label{sec:Sensing_as_a_Service Model}

\begin{figure}[t]
 \centering
 \includegraphics[scale=.26]{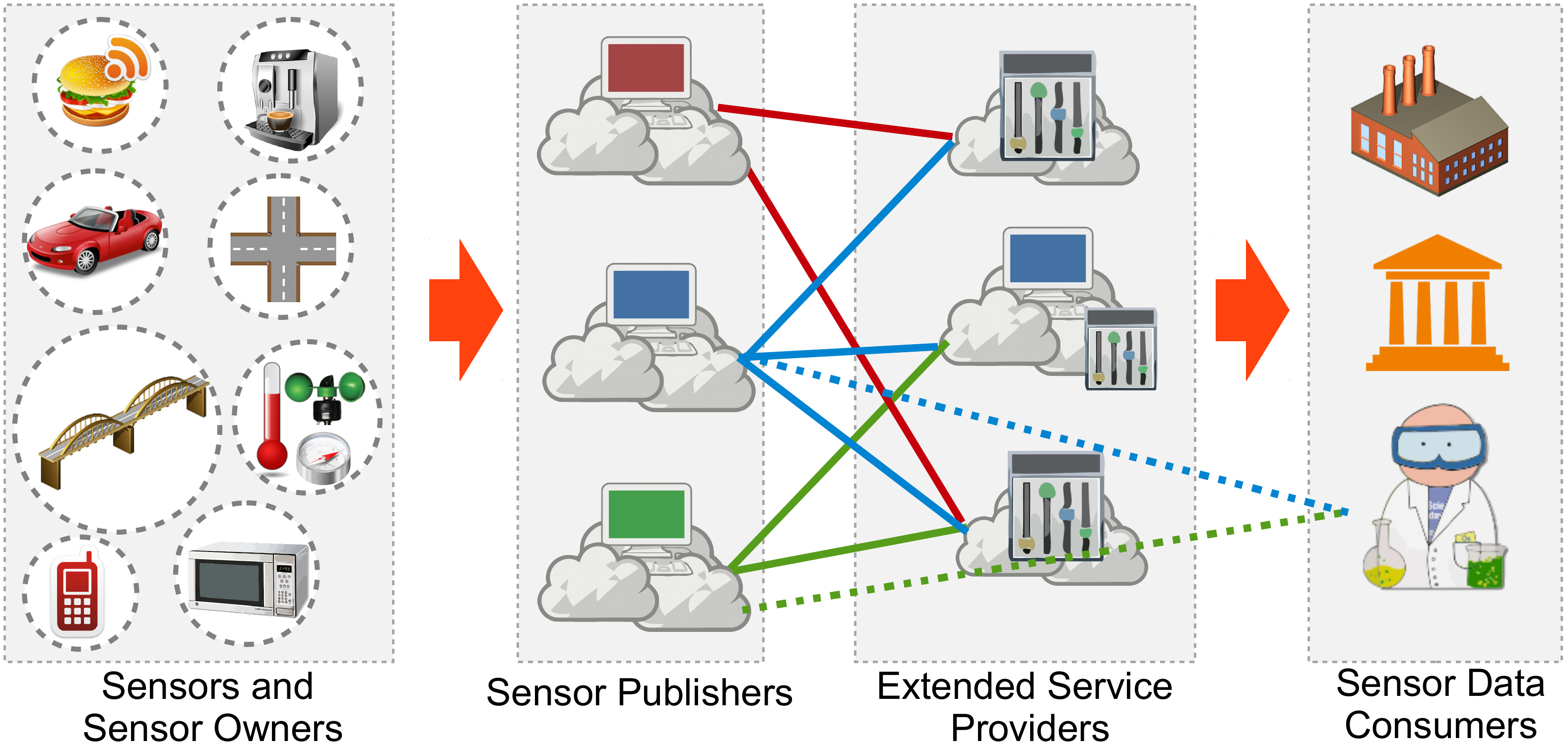}
 \caption{The sensing as a service model}
 \label{Figure:SENaas_Model}	
\vspace{-0.63cm}	
\end{figure}

In this section, we briefly discuss the sensing as a service model, how it works, and the architecture in general. Sensing as a service is a general concept that can be defined and implemented in many different ways. We propose an architectural design to maximize the value creation and sustainability of sensing as a service model as well as the Internet of Things paradigm as depicted in Figure \ref{Figure:SENaas_Model}. It consists of four conceptual layers: 1) \textit{sensors and sensor owners}, 2) \textit{sensor publishers}, 3) \textit{extended service providers}, and 4) \textit{sensor data consumers}.

\begin{figure}[t]
 \centering
 \includegraphics[scale=.34]{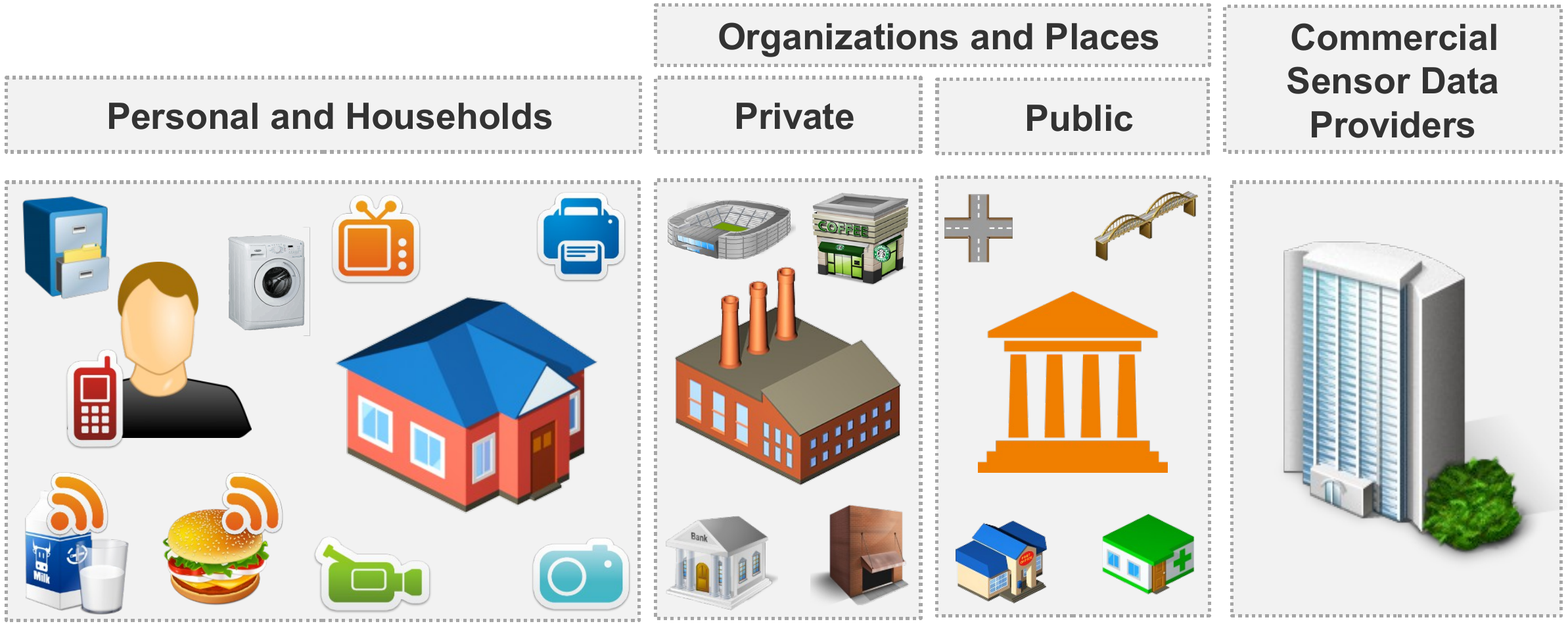}
 \caption{Sensor classification scheme based on ownership}
 \label{Figure:Classification_based_on_Sensor_Ownership}	
\vspace{-0.60cm}	
\end{figure}

\textbf{Sensors and Sensor Owners Layer:} This layer consists of sensors and sensor owners. A sensor is a device that detects, measures or sense a physical phenomenon such as humidity, temperature, etc. \cite{P009}. Multiple sensors can be attached to an object or device. For example, microwaves or coffee machines may have sensors that can be used to detect events (e.g. the number of times it is used per day and related context information). Such information can be used to understand user behaviour and user preferences more accurately. A road may have sensors that can detect the weather and traffic conditions. Today, large varieties of different sensors are available. They are capable of measuring a broad range of phenomena \cite{ZMP007}. Further, they have the capability to send  sensor data to the cloud. On the other hand, a sensor owner has the ownership of a specific sensor at a given time. Ownership may change over time. We classify sensors into four categories based on ownership as depicted in Figure \ref{Figure:Classification_based_on_Sensor_Ownership}: \textit{personal and household}, \textit{private organizations / places}, \textit{public organizations / places}, and \textit{commercial sensor data providers}. 

\vspace{-10pt}
\begin{noindlist}
\item All  personal items, such as mobile phones, wrist watches, spectacles, laptops, soft drinks, food items and household items, such as televisions, cameras, microwaves, washing machines belong to the personal and household category. In simple terms, all  items (and also all sensors) not own by private or public organizations belong to this category. We expect that all of these items (also called things, objects, and devices) would be equipped with sensors in the future.

\item The private organizations and places category consists of all  items own by private organizations. The same items we listed under personal and household category can be listed under here as well depending on the ownership. If a private company owns a coffee machine and a microwave which cannot be attributed to a single person, then those items can be listed under this category. Therefore, the private business organization has the right to take the decision whether to publish the sensors attached to those items to the cloud or not. As another example, if a private business organization owns a sport complex or a hospital, all the sensors deployed in those properties are also owned by them. When a company manufactures and sells a product that comprises sensors, the ownership get transferred to that customer. As a result, a customer will decide whether to publish those sensors in the cloud or not. 


\item The public organizations and places category is similar to the private organizations and places category we discussed above. However, this category also includes public infrastructure such as bridges, roads, parks, etc. All the sensors deployed by the government will be published in the cloud depending on government policies.

\item Commercial sensor data providers are business entities who deploy and manage sensors by themselves by keeping ownership. They earn by publishing the sensors and sensor data they own through sensor publishers. They may deploy sensors across all places such as households, private and public owned properties depending on demand and strategic value by also complying with legal terms. Mostly, they will focus on public and private places. They will also make a payment to the property owner as an exchange for giving permissions for sensor deployment. For example, commercial sensor data provider may deploy sensors in a children's park owned by state government (under government permission) to detect motion and measure the micro climate (e.g. temperature, humidity, wind speed, wind direction). Such monitoring allows to detect and predict potential crowd movements. The sensor data that can be used to predict such movements can be sold to sensor data consumers such as mobile stall businesses and children's product retailers who may be located in nearby areas.

\end{noindlist}

A sensor owner makes the final decision on whether to publish the sensors he owns in the cloud or not. If the owner decides not to publish, no sensor publisher would be able to get access to those sensors which significantly protect the security and privacy of the sensor owner. If the sensor owner decides to publish the sensors he owns, he needs to register himself with a sensor publisher. Sensor owners can define restrictions and conditions such as who can request permission and the expected return (offer). It is important to note that each sensor may send data to a different SP in the cloud (similar as we use Internet service providers). However, a single sensor only sends data to a single SP. Data will be shared between SPs if necessary depending on consumer requirements. 


\textbf{Sensor Publishers Layer:} This layer consists of sensor publishers (SP). The main responsibility of a sensor publisher is to detect available sensors, communicate with the sensor owners, and get  permission to publish the sensors in the cloud. Sensor publishers are separate business entities. When a sensor owner registers a specific sensor, SP collects information about the sensor availability, owner preferences and restriction, expected return, etc. All this information needs to be published in the cloud. Once the registration is done, a SP waits until a sensor consumer makes a request. When a SP receives such a request, it forwards all the details including the offer to the corresponding sensor owner(s) to accept or reject. If the sensor owner accepts the offer, the corresponding sensor data consumer will be able to acquire  data from that sensor through the SP during the period mentioned in the agreement (offer). The same interaction explained above can take place between SPs and ESPs. SPs entirely depend on the payments (e.g. commission) receives from sensor owners, sensor data consumers or both. \textit{Xively} (xively.com) is a public cloud for the IoT that simplifies and accelerates the creation, deployment, and management, of sensors in scalable manner. Further, it allows sharing sensor data with peers. The \textit{OpenIoT} project (openiot.eu)  an open source middleware  focuses on offering utility-based sensing services.



\textbf{Extended Service Providers Layer:} This layer consists of extended service providers (ESP). This layer can be considered as the most intelligent among all the four layers which embed the intelligence to the entire service model. The services provided by ESPs can be varied widely from one provider to another. However, there are some fundamental characteristics of ESPs. To become an ESP, they have to provide value added services to the sensor data consumers. However, in some instances a single business entity can perform both sensor publisher and extended service provider roles. Each SP has access (only) to the sensors which are registered with it. When a sensor data consumer needs sensor data from multiple sensors where each sensor has been registered with different SPs, ESPs can be used to acquire data easily. ESPs communicate with multiple SPs regarding sensor data acquisition on behalf of the sensor data consumer. The ESPs depend on the payments (e.g. commission) similar to SPs. ESPs receive payments for the value added service they provided to their customers (i.e. sensor data consumers). An example value added service can be selecting sensors based on customer's requirements \cite{ZMP004}. Customers will provide their requirements in high-level (e.g. measure environmental pollution in Canberra) instead of selecting the sensors by themselves. In return, ESP will select the appropriate sensors (e.g. pH, temperature, humidity, CO$_{2}$, etc.) located in Canberra. Pinto et al. \cite{P666} have proposed  an architectural approach for telecoms to take advantage of machine-to-machine markets in the IoT domain. It explains the opportunities  business can address by providing services related to connectivity management, data management, and service provisioning.

\textbf{Sensor Data Consumers Layer:} This layer consists of sensor data consumers. All the sensor data consumers need to register themselves and obtain a valid digital certificate from an authority in order to consume sensor data. Some of the major sensor data consumers would be governments, business organizations, academic institutions, and scientific research communities. Sensor data consumers do not directly communicate with sensors or sensor owners. All the communication and  transactions need to be done through either SPs or ESPs. If a sensor consumer has the required technical capability, they can directly acquire data from sensor publishers. However, this could be very challenging. For example, selecting which sensors to use out of billions of sensors available could be an overwhelming task \cite{ZMP006}. Further, sensor data consumers may need to communicate with multiple sensor publishers to acquire the required data. However, the cost of sensor data acquisition would be lower as they are not required to pay for ESPs' value added services. Scientific research communities may be interested in such methods. The sensor consumers with less technical capabilities and expertise can acquire required sensor data through ESPs where most of the difficult tasks such as combining sensor data from multiple sensor publishers and selecting appropriate sensors based on the consumer requirements are handled. Further, sensor consumers can register their interests with both SPs, and ESPs. For example, they can express their interest by using a number of constraints. A coffee manufacture who expects to starts its business in Canberra may be interested to access the sensor data produced by coffee machines located in Canberra for a fee. Depending on the expression of interest, ESPs/SPs will notify the coffee manufacturer when a matching deal is available. In simple terms, sensor owners define what they are expecting as return for the sensor data from one end of the Sensing as a service model. On the other end, sensor consumers define what kind of sensor data they want and how much are they willing to pay (offer). SPs and ESPs are platforms that enable these transactions (deals) to take place. The sensing as a service model shares some characteristics of an auction \cite{P671}.

%

\section{Action-based IoT Market Place}
\label{sec:The Future}

\begin{figure*}[t]
 \centering
 \includegraphics[scale=.58]{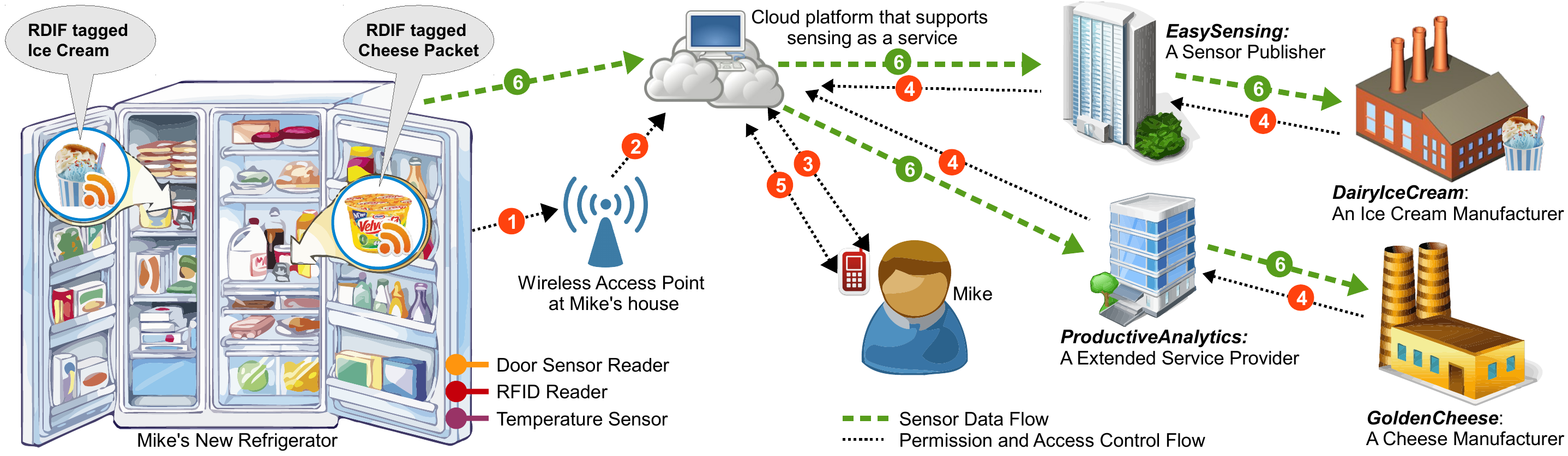}
 \caption{A futuristic scenario that explains the interactions in sensing as a service model in trading-based environment. This is not a typical smart home scenario where refrigerator tells the users what is inside, what need to be shopped or what kind of recipes can be prepared for dinner.}
 \label{Figure:UseCase}	
\vspace{-0.60cm}	
\end{figure*}

In this section, we use an example scenario, as depicted in Figure \ref{Figure:UseCase}, from smart home domain to explain the action-based IoT market place. Our intention is to highlight the interactions between different parties explained earlier in section.

\textit{Mike} bought a new refrigerator for his new home. He brought it home and plugged it to the power. The fridge automatically identifies the availability of Wi-Fi in the house as shown in step (1). Further, the refrigerator communicates with a sensor publisher and informs about its presence by providing information such as the available sensors (e.g. RFID reader, temperature, door sensors) as shown in  step (2). 
Next, in step (3), the SP communicates with Mike to check whether he likes to publish the sensors attached to the refrigerator in the cloud (step 3). We assume that \textit{Mike} has already registered with the SP in a previous transaction. \textit{Mike} is allowed to define which sensors to publish, what kind of consumers are allowed to bid, and what kind of return (fee or any other offer) is expected. Later, \textit{Mike} receives an email from a company called \textit{DairyIceCream} (via a SP called \textit{EasySensing}), an ice cream manufacturer, with an offer as shown in  step (4). \textit{DairyIceCream} is interested to have access to the RFID reader and the door sensor attached to the freezer in  \textit{Mike's} refrigerator. As a return, \textit{DairyIceCream} is willing to offer either 3\% discount on every product purchased from \textit{DairyIceCream} or a monthly fee of \$2. As \textit{Mike} likes \textit{DairyIceCream} products, he agrees to the 3\% discount offer instead of the monthly fee as shown in  step (5). A week later, \textit{Mike} receives an email from a company called \textit{ProductiveAnalytics} which has been sent on behalf of the \textit{GoldenCheese} company, a cheese manufacturer, with an similar offer. This request  also comes through \textit{EasySensing}. However, the offer is either 4\% discount on every product purchase by \textit{GoldenCheese} or a monthly fee of \$1. As \textit{Mike} does not like \textit{GoldenCheese} products, he decides to accept the monthly fee option. 


\textbf{Scenario from model perspective:} In Section \ref{sec:Sensing_as_a_Service Model}, we explained the sensing as a service model in a generic perspective and now we describe it from the above mentioned scenario perspective. In the scenario, \textit{Mike} is the sensor owner. Therefore, he and his sensors represent the \textit{sensors and sensor owners} layer. Further, in ownership categorization, \textit{Mike} represents the \textit{Personal and households} scheme. Both the \textit{DairyIceCream} and \textit{GoldenCheese} companies represent the \textit{sensor data consumers} layer. \textit{EasySensing} is a SP who enables the communication and transactions between \textit{Mike} and the \textit{DairyIceCream}. \textit{EasySensing} is responsible for matching the sensor owners expectations  with the requirements of sensor data consumers. \textit{DairyIceCream} retrieves the data from \textit{EasySensing} directly and conducts the data analysis with the help of in-house experts. \textit{ProductiveAnalytics} is an ESP who works on behalf of \textit{GoldenCheese}. \textit{GoldenCheese} has hired \textit{ProductiveAnalytics} to perform the data analysis as they do not have the required technical skills within the company. \textit{ProductiveAnalytics} collects data by handling all the deals and transaction with the sensor owners though their partner SPs.

\section{Value Creation and Social Impact}
\label{sec:Value_Creation}

So far we discussed how trading-based sensing as a service model works in architectural  point of view.  In this section, we explain how everything fits together in order to create value for the consumers and its impact towards society.


 \textbf{\textit{Win-Win situation}:} As we mentioned before, one of the main goal of IoT is to create \textit{`a better world for human beings'}, where objects around us know what we like, what we want, and what we need and act accordingly without explicit instructions \cite{P040}. Most of the IoT solutions \cite{P596} in the market place are designed to step towards achieving this objective. Especially, when ageing population is increasing, such solutions can make a significant impact to improve the well-being and quality of life (e.g. elderly care). However, cost of such solutions have put many people away from adopting these solutions. Action-based IoT market place driven by sensing as a service paradigm has the capability to address this problem. The proposed model can help the consumer to earn back the additional costs that they may need incur due to adoption of smart IoT solutions. This allows consumers to experience the comfort offered by IoT solutions without additional financial overheads. We believe such encouragement can motivate consumers to adopt IoT solutions much faster and sustainable way. The extra earning that consumers  may gain through participating in trading-based sensing as a service model will motivate them to continuously maintain and (repair / replace) IoT hardware infrastructure around them (in households). The potential earning capability will also motivate household occupants to purchase smart  devices (e.g. refrigerator, microwaves, sensor enabled furniture, air condition and lighting systems, etc.) that supports IoT as well as trading-based sensing as a service model even for higher prices in comparison to traditional devices.

The above logical explanation stands true for all other domains as well. For example, a farmer can deploy IoT solutions on his filed to monitor frost events, diseases detection and so on and still cover the cost of deploying such solutions via participating in the auction-like trading model. The proposed sensing as a service model not only motivates individuals but also corporations to actively engage  in deploying IoT solutions. The model creates a win-win situation for all the parties involved. Based on the scenario we presented in Section \ref{sec:The Future}, \textit{Mike} (sensor owners' perspective) is getting a return (a valuable offer). Additionally, he also receives the comfort that the IoT solutions typically provides (e.g. refrigerator tells the users what is inside, what need to be shopped or what kind of recipes can be prepared for dinner.). In \textit{DairyIceCream}  perspective, now they have real-time data about product consumer behaviour (e.g. when \textit{Mike} eats ice cream, how frequent, whether \textit{Mike} use substitutions and so on). Therefore, \textit{DairyIceCream} is no longer required to conduct manual surveys and market analyses.


 \textbf{\textit{Sharing and reusing}:} In traditional methods, each party (group or person) who wants to collect sensor data needs to visit the field and deploy the sensors manually by themselves. Further, there is no easy way to share  sensor data collected by one party with others. Sensing as a service is a model that stimulates by concept of sharing. In simple terms, if someone has already deployed the sensors, others can have access to them by paying a fee to the sensor owner. One of the major arguments that could arise regarding sensing as a service model is that \textit{``How to convince a manufacturer to embed sensors and communication capabilities into devices we use in everyday life (e.g refrigerator in the use-case presented in Section \ref{sec:The Future})''}. This question can be answered in two different perspectives.  First, IoT envisions to have sensor embedded into objects around us. The goal of IoT is to allow devices to communicate with each other. Naturally, such a goal forces next generation devices to be embedded with rich sensing and communication capabilities. Therefore, the motivation is given to the manufacturers not by the sensing as a service model but the vision of IoT.  The sensing as a service model is designed to provide incentives to users which motivate them to purchase next generation devices that supports both IoT envisioned interactions as well as the sensing as a service model. The additional cost that contributes to increase the prices of the devices (due to embedding rich sensing and communication capabilities) can be easily covered by participating in the sensing as a service model itself. Even today, state of the art devices such as refrigerators and televisions comprise communication and sensing capabilities. Due to the shared and collaborative nature,  data acquisition cost will be reduced significantly. Such a sustainable business model stimulates more and more sensor deployments. Further, technological advances and higher demands allow to produce sensors in mass volumes by reducing the cost per unit. 


 \textbf{\textit{Collect data previously unavailable}:} This model allows to collect sensor data which is impossible to collect using traditional non-collaborative methods. This business model promotes and stimulates the sensor deployments by companies at commercial level. As we explained earlier in Section \ref{sec:Sensing_as_a_Service Model}, dedicated business entities will deploy sensors in public places such as parks and bridges so  government authorities can have access to those sensors by paying only for the data they need in real-time or archived. Today business entities spend substantial amount of money to conduct market analyses and consumer surveys. A sample of 1,000 respondents, which would give a statistical accuracy of +/-3.1\% costs around \$8,000 \cite{P632}. Recently, different third party companies started offering consumer surveys on behalf of businesses. One such solution is Google Consumer Surveys. It allows businesses to target user groups with specific criteria and conduct the survey. Currently, one user response (for one question) cost around \$0.10, 1/10th of the cost of similar quality research conduct using traditional methods. Even though such approaches have reduced the cost of surveys, they still cost substantial amount and have deficiencies such as latency, inaccuracies, and so on. In the sensing as a service model, all the data is directly coming from the sensor without user intervention. This also helps to reduce the cost of data acquisition. Due to privacy concerns it is important to anonymise the sensor data collected. 
 



\section{Community View and Discussion}
\label{sec:Discussion}

In this section, we discuss results of two surveys we conducted in order  to find out the public opinion towards the sensing as a service model.

\begin{figure}[h]
 \centering
 \includegraphics[scale=0.9]{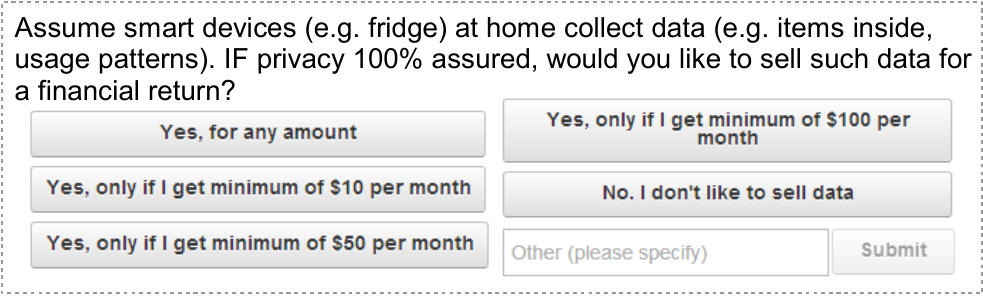}
\vspace{-0.33cm}	
 \caption{Survey 1: (1 question / 1000 respondents)}
 \label{Figure:Google_Consumer_Surveys}	
\vspace{-0.33cm}	
\end{figure}

\begin{figure}[t]
 \centering
 \includegraphics[scale=0.4]{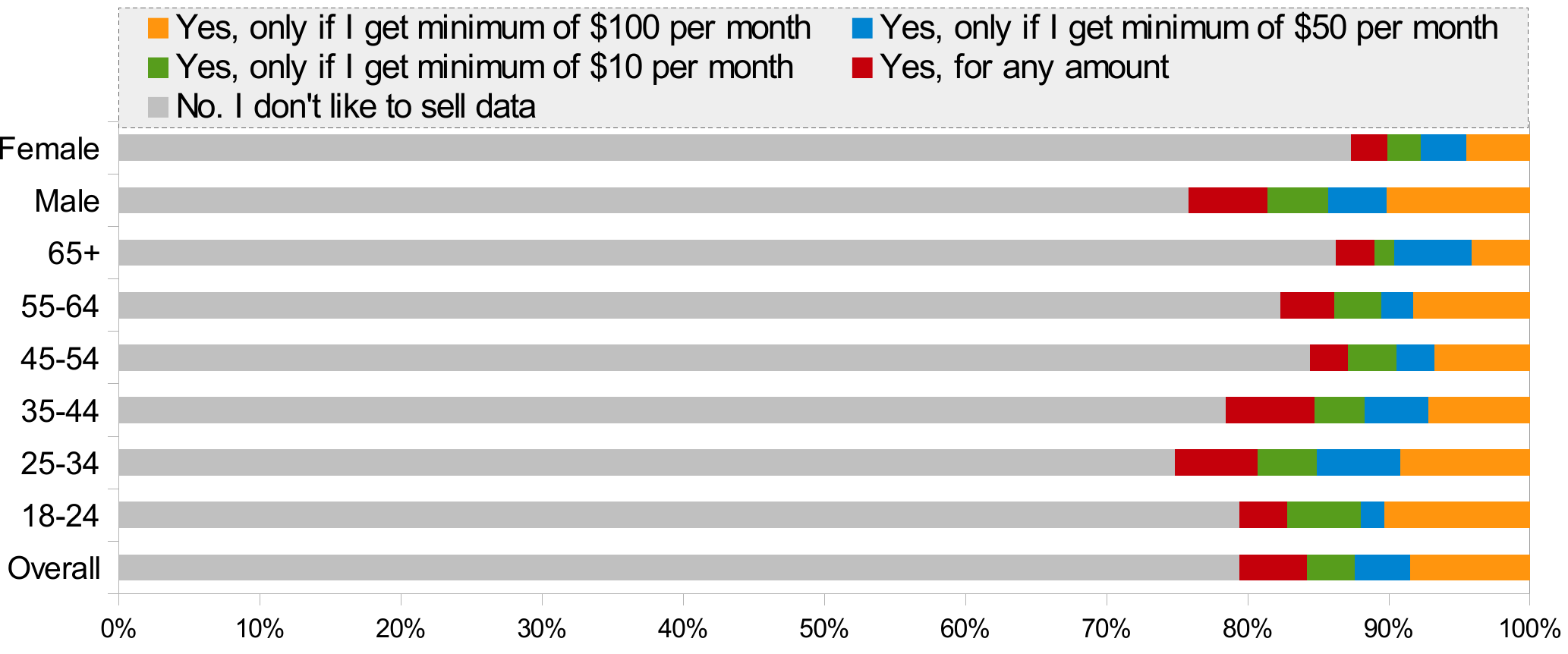}
\vspace{-0.33cm}	
 \caption{Survey 1 Results}
 \label{Figure:Google_Consumer_Surveys_Results1}	
\vspace{-0.72cm}	
\end{figure}

\textbf{Survey 1 overview:}  We conducted survey 1 through \textit{Google Consumer Surveys}\footnote{http://www.google.com/insights/consumersurveys/} platform where we collected 1000 responses. The \textit{Google Consumer Surveys} is a single question survey as  depicted in Figure \ref{Figure:Google_Consumer_Surveys}. The question was targeted on United States general population. The objective of this survey is to find out the public opinion towards \textit{`selling sensor data generated by the smart devices at home for a financial reward'}. Due to survey platform limitations, we could not provide detailed explanations to the respondents that explain the context (i.e. Internet of Things, sensing as a service model, trading). The results are presented in Figure \ref{Figure:Google_Consumer_Surveys_Results1}.

\textbf{Survey 1 Discussion:} The majority of responses are negative (79\%). According the result depicted in Figure \ref{Figure:Google_Consumer_Surveys_Results1}, young respondents are more positive about the sensing as a service model than older respondents. Further, males respondents are also positive towards the model in comparison to female respondents. We believe reason for negative responses is due to lack of understanding. This is clearly evident when survey 2 is analysed. After analysing the opinions that respondents are submitted through open text box, it is realized that privacy has been the main concern. Even though, in the question we explicitly requested to assume 100\%, it has not been well received by the respondents. Finally, it is important to note that 21\% of the respondents have positively responded towards the proposed model despite the lack of background knowledge (IoT and sensing as a service) that we have been provided in the question.

\begin{figure}[b]
 \centering
 \vspace{-0.43cm}
 \includegraphics[scale=0.42]{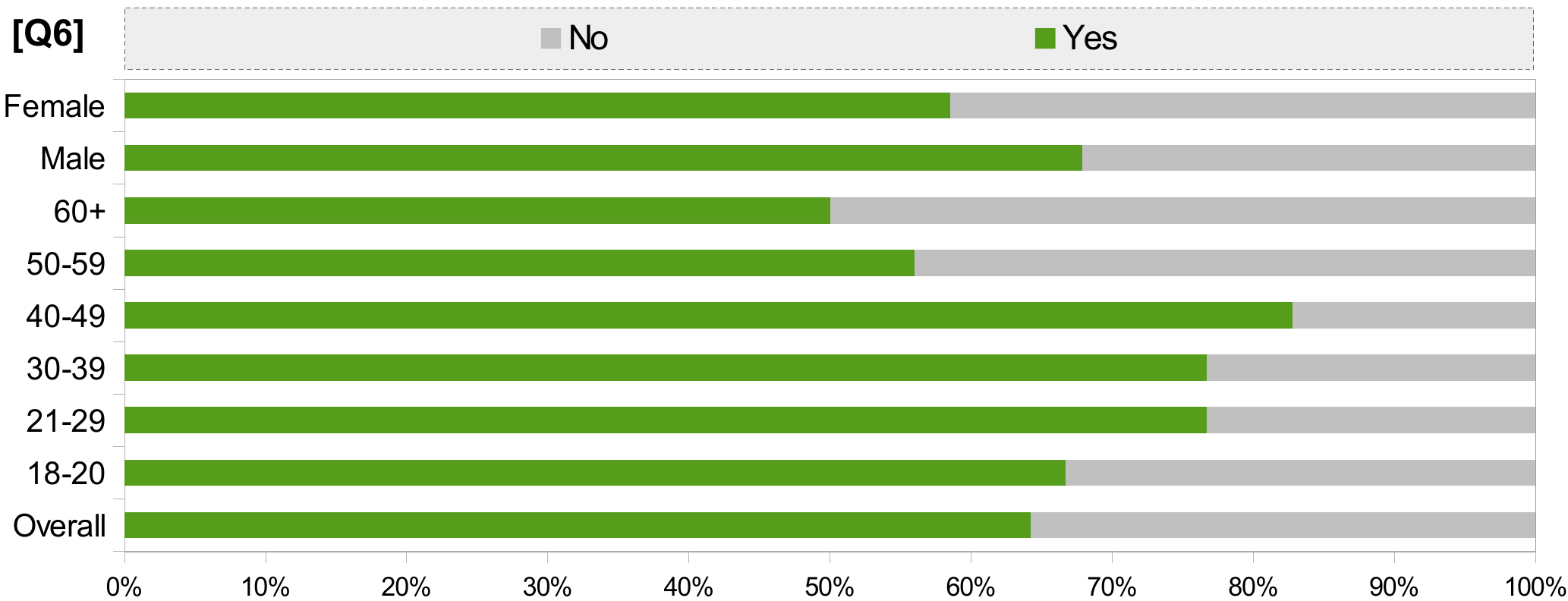}
 \caption{Survey 2 Results (9 questions / 137 responses)}
 \label{Figure:SurveyMonkey_Surveys_Results2}	
\vspace{-0.52cm}	
\end{figure}

\textbf{Survey 2 overview:}  The survey 2 has been administrated using  \textit{SurveyMonkey}\footnote{https://www.surveymonkey.com/} platform. It was a multi-question survey that contained total of nine questions (including five demographics questions i.e. country, gender, age, household income, IT knowledge / familiarity). We collected total of 137 responses. The questions we asked in this survey 2 are as follows. The survey is available on-line (goo.gl/N3h4mq). The results are presented in Figure \ref{Figure:SurveyMonkey_Surveys_Results2}.

\begin{figure}[h]
 \centering
 \includegraphics[scale=0.9]{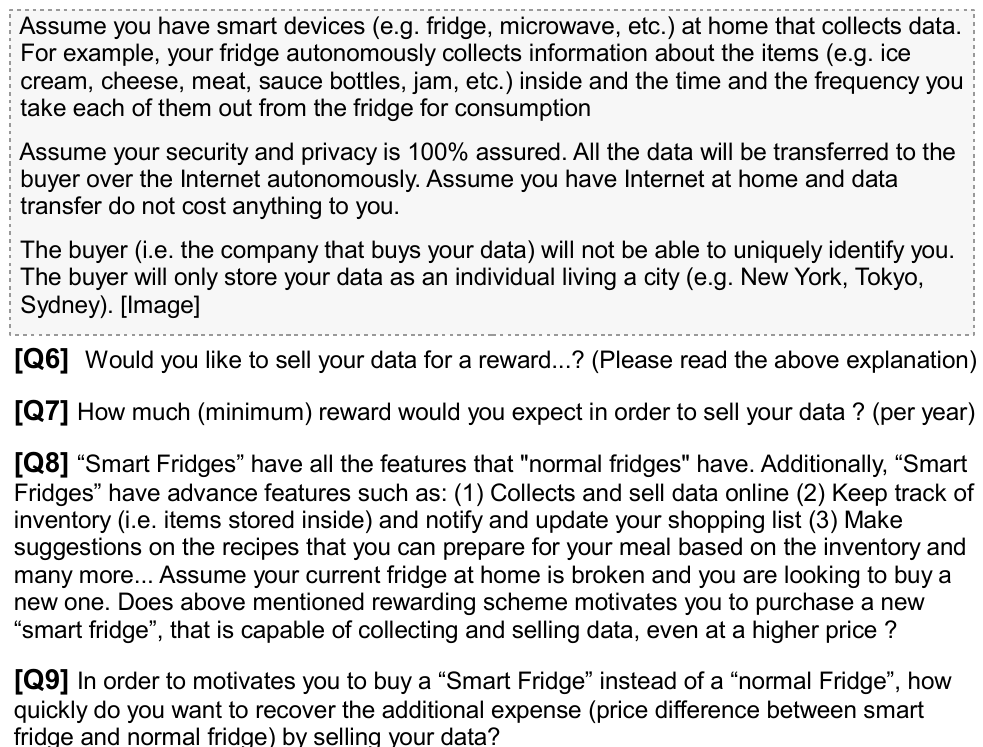}
 \caption{Survey 2: (9 questions / 137 respondents)}
 \label{Figure:SurveyMonkey_Surveys}	
\vspace{-0.44cm}	
\end{figure}

\begin{figure}[t]
 \centering
 \includegraphics[scale=0.58]{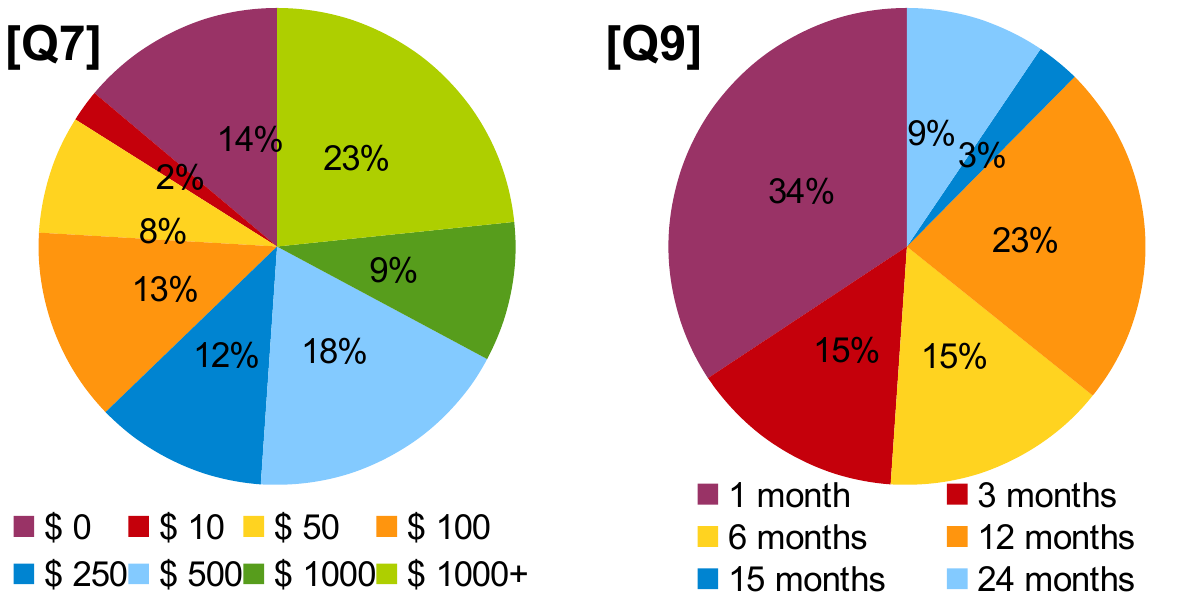}
 \caption{Survey 2: Results for [Q7] and [Q9]}
 \label{Figure:SurveyMonkey_Surveys_Results3}	
\vspace{-0.60cm}	
\end{figure}

\textbf{Survey 2 Discussion:} In contrast to the survey 1 results, majority of the responses (64\%) in survey 2 are in favour of trading-based sensing as a service model. We believe that the positivity is due to the detailed explanation we provided so the respondents have better idea about the model the its context. According to the Figure \ref{Figure:SurveyMonkey_Surveys_Results2},  young respondents are more positive about the proposed model than older respondents. Further, males respondents are also positive towards the model in comparison to female respondents. These two observations validate the results of survey 1 as well. The responses for Q7 and Q9 are summarized in Figure \ref{Figure:SurveyMonkey_Surveys_Results3}. According to the results, 67\% of the respondents expect less than US\$500 per year.

\begin{figure}[t]
 \centering
 \includegraphics[scale=0.4]{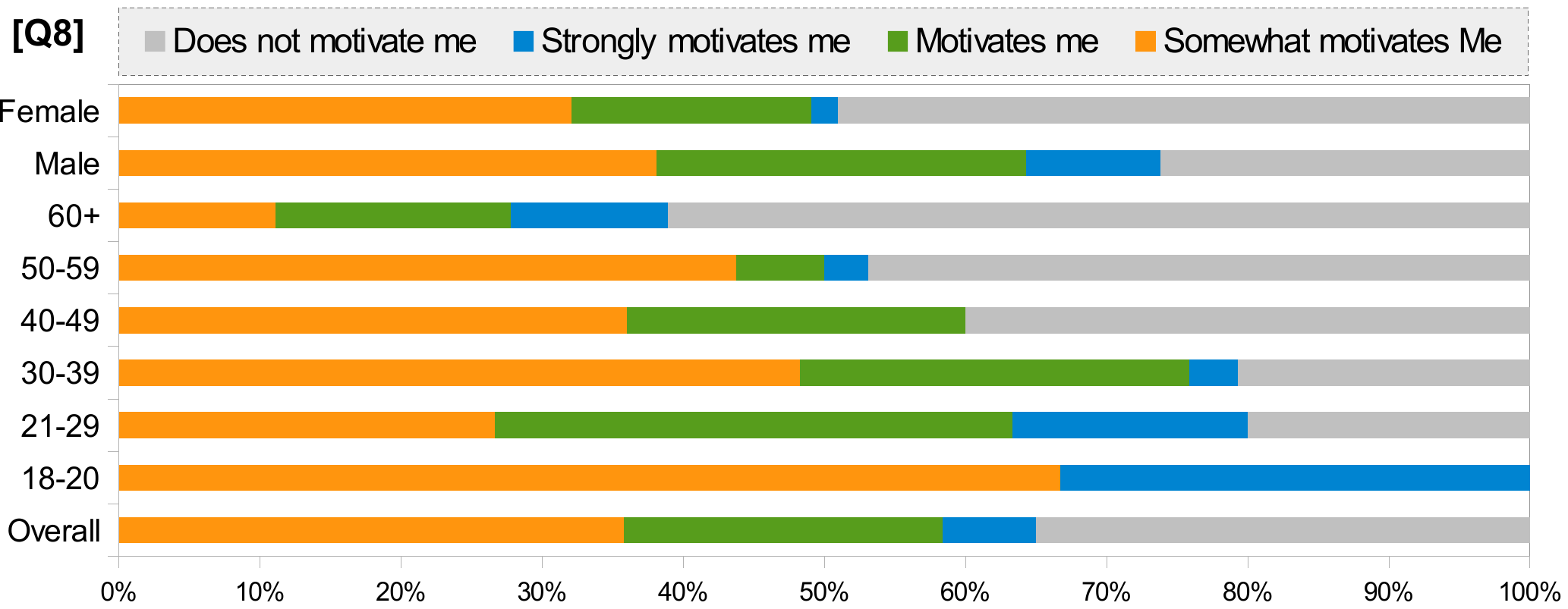}
\vspace{-0.23cm}	
 \caption{Survey 2: Results for [Q8]}
 \label{Figure:SurveyMonkey_Surveys_Results4}	
\vspace{-0.72cm}	
\end{figure}

The value of data generated (e.g. supply chain management, usage prediction, waste reduction due to smart appliances and infrastructure, consumer satisfaction and surveys etc. $\times$ number of organization interest in data $\times$ 365/24/7 data access) by IoT solutions  deployed in smart household environments can exceeds the  financial value  that respondents expect. According to  \ref{Figure:SurveyMonkey_Surveys_Results3} [Q9], 66\% respondents are happy to make additional investments as long as the additional cost can be covered within 3 months. This allows manufactures to embed smart sensing and communication technologies into smart household appliances and infrastructure. Finally, Figure \ref{Figure:SurveyMonkey_Surveys_Results4} clearly shows that our proposed model motivates respondents (65\%) to purchase smart devices and adopt IoT solutions even at higher prices. Similar to the previous observations, younger respondents and male respondents have been motivated more in comparison older and female respondents.

The overall general public opinion towards the proposed model is positive when the details are explained, and privacy concerns are omitted. The central concern is the security and privacy. It is essential to note that the surveys have been conducted during a period that online privacy has been in the spotlight due to governments' surveillance programs. The disparity between the results of two surveys highlights the importance of clear and concise communication of the benefits and advantages of the model. Such communication allows to gain the trust of the consumers. When we consider all the factor, the proposed model can increase the adoption and sustainability of the IoT trading based value creation by creating a win-win situation for everyone involved. Further, the proposed model will help to overcome the challenges in adoption of home automation IoT solutions discussed in \cite{Z001}. Specially, consumers will be motivated to deploy IoT solutions  in their households by themselves as they have the opportunity to receive financial rewards.

\section{Conclusion}
\label{sec:Conclusion}

In this paper, we explored the methods that increase the adoption and sustainability of IoT solutions. We provided a comprehensive overview of the sensing as a service model and its applicability towards the Internet of Things paradigm. We proposed an business model that encourages non-technical households occupants to adopt IoT solution without bearing significant costs. Specifically, proposed model allows them to experience the comfort delivered by state of the art IoT solution where the cost of adoption can be covered with participating in market (i.e. selling data). Therefore, it creates a win-win situation for all the parties involved. Based on the result of the surveys, it is evident that the general public has a positive opinion on such a model. This will increase the adoption and the sustainability of IoT. Finally, this model will create an unprecedented amount of opportunities to build innovative value added solutions that improve the well-being and quality of life of the citizens in smart cities.




\textbf{Acknowledgements:} Authors acknowledge support from OpenIoT Project, FP7-ICT-2011-7-287305-OpenIoT.





%
  \bibliography{Bibliography}
  \bibliographystyle{abbrv}

\end{document}